\pdfoutput=1
\documentclass[hyphens]{article}
\usepackage{fullpage,mathptmx,amsmath,amsthm}
\usepackage{microtype}
\usepackage{graphicx}
\usepackage{xspace}
\usepackage{amsmath}
\usepackage{amssymb}
\usepackage{paralist}
\usepackage{varwidth}
\usepackage{cite}
\usepackage{hyperref}
\usepackage{url}
\PassOptionsToPackage{hyphens}{url}

\usepackage{acronym}

\acrodef{meb}[MEB]{minimum enclosing ball}

\newcommand{\eat}[1]{}
\newcommand{\poly}{\text{poly}}
\newcommand{\field}{\mathbb{F}}
\newcommand{\eps}{\varepsilon}

\newcommand{\bx}{\mathbf{x}}
\newcommand{\by}{\mathbf{y}}
\newcommand{\bw}{\mathbf{w}}
\newcommand{\bz}{\mathbf{z}}

\newcommand{\pq}{\textsf{PointQuery}\xspace}
\newcommand{\finger}{\textsf{finger}\xspace}
\newcommand{\range}{\textsf{RangeCount}\xspace}

\newcommand{\IMM}{\textsf{MatrixMultiplication}\xspace}
\newcommand{\reals}{\ensuremath{\mathbb{R}}}

\renewcommand{\v}[1]{\ensuremath{\mathbf{#1}}}
\newcommand{\ind}{\ensuremath{\mathbb{I}}}

\newtheorem{theorem}{Theorem}[section]
\newtheorem{lemma}[theorem]{Lemma}

\newtheorem{definition}[theorem]{Definition}

\date{}
\begin{document}
\title{Streaming Verification in Data Analysis\thanks{This research was
    supported in part by the National Science Fondation under grants
    BIGDATA-1251049, CPS-1035565 and CCF-1115677}} 

\author{Samira Daruki\thanks{School of Computing, University of Utah} \and  Justin Thaler\thanks{Yahoo Labs} \and Suresh Venkatasubramanian\thanks{School of Computing, University of Utah}}

\maketitle

\begin{abstract}
  Streaming interactive proofs (SIPs) are a framework to
  reason about outsourced computation, where a data owner (the verifier) 
  outsources a computation to the cloud (the prover), but wishes to verify
  the correctness of the solution provided by the cloud service.
  In this paper we present streaming interactive proofs for problems in data
  analysis. We present protocols for clustering and shape fitting problems, as
  well as an improved protocol for rectangular matrix multiplication. The latter
 can in turn be used to verify $k$ \emph{eigenvectors} of a (streamed)
  $n \times n$ matrix. 

  In general our solutions use polylogarithmic rounds of communication and polylogarithmic total
  communication and verifier space. For special cases (when optimality
  certificates can be verified easily), we present constant round protocols with
  similar costs. For rectangular matrix multiplication and eigenvector verification, 
  our protocols work in the more restricted annotated data streaming
  model, and use sublinear (but
 not polylogarithmic) communication. 
\end{abstract}


\section{Introduction}
\label{sec:introduction}
There are now many third party ``cloud'' services (from companies like Amazon, Google and Microsoft) that can perform intensive computational tasks on large data. 
Computing effort is split between a computationally weak ``client'' who  owns the data and wishes to solve a desired task, and a ``server'' consisting of a cluster of computing nodes that performs the computation. 

In this setting, how does a client verify that a computation has been performed correctly? The client here will have limited (streaming) access to the data, as well as limited ability to talk to the server (measured by the amount of communication and rounds). Recently, there has been  renewed interest in studying interactive verification with extremely limited \emph{sublinear space} (or streaming) verifiers. Such  \emph{streaming interactive proofs} (SIPs) have been developed for classic problems in streaming, like frequency moment estimation and related graph problems. 


\subsection{Our Contributions}

We initiate a study of streaming interactive proofs for problems in data analysis. In what follows, we will refer to both SIPs and \emph{annotated streaming protocols} which are a variant of SIPs (we discuss the models and their differences in Section \ref{sec:background}). 

\medskip
\noindent \textbf{Matrix Analysis.}  We present an annotated data streaming protocol (Section~\ref{sec:verify-matr-eigenstr}) for rectangular matrix multiplication over any field $\field$. Specifically, given input matrices $A \in \field^{k \times n}$ and $B \in \field^{n \times k'}$,
our protocol computes their product, using communication cost $k \cdot k' \cdot h \log |\field|$ and space cost $v \log |\field|$, for any desired pair of positive integers $h, v$ satisfying $h \cdot v \geq n$. This improves on prior work \cite{cormode2013streaming} by a factor of $k$ in the space cost, and we prove that this tradeoff is optimal up to a factor of $\tilde{O}\left(\min\left(k, k'\right)\right)$. 
The rectangular matrix multiplication protocol can in turn be used to verify $k$ (approximate) eigenvectors of an $n \times n$ integer matrix $A$.\footnote{We cannot in general verify that the provided vectors are exact eigenvectors due to precision issues. Section \ref{sec:eigen} has details.}

\medskip
\noindent \textbf{Shape Analysis.} We present a number of protocols for shape fitting and clustering problems.
\begin{inparaenum}[(i)]
\item We give \emph{3-message} SIPs that can verify a \ac{meb} and the width of a point set \emph{exactly}  with polylogarithmic space and communication costs. Note that the \ac{meb} cannot be approximated to better than a constant factor by a streaming algorithm with space even polynomial in the dimension \cite{agarwal2010streaming}: 
we show that the streaming hardness of the \ac{meb} problem holds even when the points are chosen from a discrete cube: this is important because our interactive proofs require discrete input (Section \ref{sec:verify-optim-few}).
\item We present polylogarithmic round protocols with polylogarithmic communication and verifier space for verifying optimal $k$-centers and $k$-slabs in Euclidean space (note that computing the \ac{meb} and width of a point set correspond to the $1$-center and $1$-slab problems respectively) (Section~\ref{sec:verify-clust-probl}).
\item We also show a simple 3-message protocol for verifying a 2-approximation to the $k$-center in a metric space, via simple adaptation of the Gonzalez 2-approximation for $k$-center (Section \ref{sec:verify-optim-few}).
\end{inparaenum}

\paragraph{Technical Overview.}
\label{sec:technical-overview}
In our annotated data streaming protocol for matrix multiplication, we first observe that multiplying a $k \times n$ matrix $A$ with an $n \times k'$ matrix $B$ is equivalent to 
performing $k'$ \emph{matrix-vector} multiplications, one for each column of $B$. But rather than naively implement $k'$ matrix-vector verification protocols \cite{cormode2013streaming}, we exploit the fact that the $k'$ matrix-vector multiplications are not independent, because the matrix $A$ is held fixed in all of them.
This leads to an improved subroutine for rectangular matrix multiplication that in turn allows us to verify eigenvectors of a matrix. 

For the $k$-center and $k$-slab problems, we must verify feasibility and optimality of a claimed solution. We verify feasibility by reducing to an instance of a Range Counting problem, for which a 2-message SIP exists \cite{chakrabartiinteractivity}. For optimality, the prover must convince the verifier that \emph{no} other feasible solution has lower cost. When $k=1$, we show that there is a \emph{sparse} witness of optimality, which the verifier can check directly using 3 messages, by reduction to Range Counting. For general $k$, we cannot produce such a witness. However, we observe that the ``for-all'' constraint on feasible solutions (that they all be costlier than the claimed solution) can be expressed as a sum over all solutions of potentially lower cost. Choosing a cost-based ordering of solutions converts  this into a partial sum over a prefix of the ordered set of solutions. Our main tool is a way to verify such a sum in general, using polylogarithmically many messages, \emph{even when the relevant prefix is only known after the stream has passed}. 

We note that while \emph{core sets} are a natural witness for a property of a point set, they cannot always be computed by a streaming algorithm, nor is it clear that a claim of being a core set is easily verified. For the problems considered here, these issues preclude the use of a ``simple'' core set, requiring a more complex interactive protocol.



\subsection{Prior Work on Streaming Verification}
\label{sec:related-work}
Chakrabarti et al. \cite{chakrabarti2014annotations,chakrabarti2009annotations} introduced the notion of \emph{annotations} in data streams, whereby an all-powerful prover could provide annotations to a verifier in order to complete a stream computation. Cormode et al. \cite{cormode2011verifying} introduced the model of Streaming Interactive Proofs (SIPs), which extends the annotated data streaming model to allow for multiple rounds of interaction between the prover and verifier. 
They introduced a streaming variant of the classical sum-check protocol \cite{lfkn}, and used it to give logarithmic cost protocols for a variety of well-studied streaming problems. In subsequent works, protocols were developed in both models for graph problems and matrix-vector operations \cite{cormode2013streaming}, \emph{sparse streams} \cite{chakrabarti2013annotations}, and were implemented \cite{cormode2012practical}. Most recently, Chakrabarti et al. \cite{chakrabartiinteractivity} developed streaming interactive proofs of logarithmic cost that worked in $O(1)$ rounds, making use of an interactive protocol for the \textsc{Index} problem. Lower bounds on the cost of SIPs and their variants have also been studied \cite{babai1986complexity,chakrabarti2013annotations,chakrabartiinteractivity,klauck2011arthur, gur}. These results make use of \emph{Arthur-Merlin communication complexity} and related notions.  
There has also been work in the cryptography community on stream verification protocols that are 
secure only against cheating provers that run in polynomial time (e.g., \cite{chung, eurocrypt, vds}). 
The interested reader is referred to \cite{thalersurvey} for a more detailed overview of the literature on models for stream verification.



\section{Preliminaries}
\label{sec:background}

\textbf{Models.}  We will work in the \emph{streaming interactive proof} (SIP) model first proposed by Cormode et al. \cite{cormode2011verifying}. In this model, there are two players, the prover \textsf{P} and verifier \textsf{V}. The input consists of a \emph{stream} $\tau$ of $n$ items from some universe. 
Let $f$ be a function mapping a stream $\tau$ to any finite set $\mathcal{S}$. A $k$-message SIP for $f$ works as follows. First,
 \textsf{V} and $\textsf{P}$ read the input stream. During this phase, \textsf{V} computes some small secret state, which depends on $\tau$ and \textsf{V}'s
private randomness.
Second, \textsf{V} and \textsf{P} then exchange $k$ messages, after which \textsf{V}  outputs a value in $\mathcal{S} \cup \{\perp\}$, where $\perp$ indicates that $\textsc{V}$ is not convinced by \textsf{P}. 

Any SIP for $f$ must satisfy soundness and completeness. Completeness requires that there exists some prover strategy that causes the verifier to output $f(\tau)$ with probability
$1-\eps_c$ for $\eps_c \leq 1/3$. Soundness requires that for all prover strategies, the verifier outputs a value in $\{f(\tau), \perp\}$ with probability $1-\eps_s$ for some $\eps_s \leq 1/3$. 
The values $\eps_c$ and $\eps_s$ are referred to as the completeness and soundness errors.\footnote{All of our protocols achieve perfect completeness and soundness error $1/\text{poly}(n)$.} 

\paragraph{Annotated Data Streams.} The \emph{annotated data streaming model} of Chakrabarti et al. \cite{chakrabarti2009annotations} essentially corresponds to one-message SIPs.\footnote{While the original model allowed \textsf{P} to interleave information with the stream, most known annotated streaming protocols do not do so, and are thus 1-message SIPs.}

\paragraph*{Costs.}
\label{sec:costs}
In a SIP, the goal is to ensure that 
\textsf{V} uses sublinear space and that the protocol uses sublinear communication (number of bits exchanged between \textsf{V} and \textsf{P}) \emph{after} stream observation. We will also desire protocols in which \textsf{V} and \textsf{P} can run quickly. In our protocols, both \textsf{V} and \textsf{P} can execute the protocol in time quasilinear in the size of the input stream. 

\paragraph{Input Model.}
All of the protocols we consider can handle inputs
specified in a general data stream form. Each element of the stream is a tuple $(i, \delta)$, where each $i$ lies in a data universe $\mathcal{U}$ of size $u$, and $\delta \in \{+1, -1\}$.
Negative values of $\delta$ model deletions. The data stream implicitly defines a
frequency vector $\mathbf{a}=(a_1, \dots, a_u)$, where $a_i$
is the sum of all $\delta$ values associated with $i$ in the stream.
\paragraph{Discretization.}
\label{sec:discretization}
The protocols we employ make extensive use of finite field arithmetic. In order to apply these techniques to geometric problems, we must assume that all input points are drawn from the discretized grid $\mathcal{U}=[m]^d$ as the data universe. Importantly, the costs of our protocols will depend only logarithmically on $m$, enabling the grid to be exceedingly fine while still yielding tractable costs.

\subsection{Protocols from Prior Work} We will make use of three basic tools in our algorithms: Reed-Solomon fingerprints for testing vector equality, a two-message SIP  of Chakrabarti et al. \cite{chakrabartiinteractivity} for the \textsc{PointQuery} problem, and the streaming sum-check protocol of Cormode et al. \cite{cormode2011verifying}. We summarize the main properties of these protocols here: for more details, the reader is referred to the original papers.

\subsubsection{Fingerprinting}
\begin{theorem}[Reed-Solomon Fingerprinting]\label{fingerprint}
Suppose the input stream $\tau$ specifies two vectors $\mathbf{a}, \mathbf{a}' \in \mathbb{Z}^u$, guaranteed to satisfy  $|\mathbf{a}_i|, |\mathbf{a}'_i| \leq u$ at the end of $\tau$.
There is a streaming algorithm using $O(\log u)$ space that satisfies the following properties: (i) If $\mathbf{a} = \mathbf{a'}$, then the algorithm outputs 1 with probability 1. (ii) If $\mathbf{a} \neq \mathbf{a'}$, then the algorithm outputs 0 with probability at least $1-1/u^2$. 
\end{theorem}
\begin{proof} 
Let $\field$ be a finite field of prime order, satisfying $6 u^3 \leq |\field| \leq u^4$. We view each entry of $\mathbf{a}$ and $\mathbf{a}'$ as an element $\field$ in the natural way. 
At the start of the stream, the streaming algorithm picks an $\alpha \in \field$ at random, and computes $\finger(\mathbf{a}) = \sum_{i \in [i]} a_i \cdot \alpha^i$ and $\finger(\mathbf{a}') = \sum_{i \in [i]} a'_i \cdot \alpha^i$ with a single streaming pass
over $\tau$. The algorithm outputs 1 if and only if $\finger(\mathbf{a}) = \finger(\mathbf{a}')$. Property (i) clearly holds: if $\mathbf{a} = \mathbf{a}'$, then the algorithm outputs 1 with probability 1. 
To see that Property (ii) holds, observe that $\finger(\mathbf{a})$ and $\finger(\mathbf{a'})$ are univariate polynomial of degree at most $u$. If $\mathbf{a} \neq \mathbf{a}'$, these two polynomials are not equal. Property (ii) then follows, because any two distinct polynomials of degree at most $u$ over $\field$ can agree on at most $u$ inputs, yielding an error of at most $1/u^2$. 
\end{proof}

\subsubsection{The \pq and \range Protocols}
\label{sec:textscindex-protocol}
An instance of the \pq problem consists of a stream of updates as described above followed by a query $q \in [u]$. The goal is to compute the coordinate $\mathbf{a}_q$. For \range problem, let $(\mathcal{U}, \mathcal{R})$ be a range space and the input consist of a stream $\tau$ of elements (with size $n$) from the data universe $\mathcal{U}$ (with size $u$), followed by a range $R \in \mathcal{R}$. The goal is to verify a claim by \textsf{P} that $|R \cap \tau| = k$. 

\begin{theorem}[Chakrabarti et al. \cite{chakrabartiinteractivity}]
\label{lemma:index}
Suppose the input to \pq satisfies  $|\mathbf{a}_i| \leq \Delta$ at the end of the stream, for some known $\Delta$. Then there is a two-message SIP for \pq on an input stream with length $n$, with space and communication each bounded by $O(\log u \cdot \log(\Delta + \log u))$. For \range, there is a two-message SIP for \range with space and communication cost bounded by $O(\log (|\mathcal{R}|) \cdot \log (n \cdot |\mathcal{R}|))$. In particular, for range spaces of bounded shatter dimension $\rho$, $\log |\mathcal{R}| = \rho \log n = O(\log n)$.
 \end{theorem}



\subsubsection{Sum-Check Protocol}
\begin{figure}
{
\centering
\fbox{
\small
\begin{varwidth}{11cm}
\noindent \textbf{Input:} $\textsf{V}$ is given oracle access to a $v$-variate polynomial $g$ over finite field $\mathbb{F}$ and an $H \in \mathbb{F}$.\\
\noindent \textbf{Goal:} Determine whether $H=\sum_{(x_1, \dots, x_v) \in \{0, 1\}^v} g(x_1, \dots, x_v)$. \\
\begin{itemize}
\item In the first round, $\textsf{P}$ computes the univariate polynomial $$g_1(X_1) := \sum_{x_2, \dots,x_v \in \{0, 1\}^{v-1}} g(X_1,x_2,\dots,x_v),$$ 
and sends $g_1$ to $\textsf{V}$. $\textsf{V}$ checks that $g_1$ is a univariate polynomial of degree at most $\deg_1(g)$, and that $H=g_1(0) + g_1(1)$,
rejecting if not.

\item $\textsf{V}$ chooses a random element $r_1\in \mathbb{F}$, and sends $r_1$ to $\textsf{P}$.

\item In the $j$th round, for $1 < j < v $, $\textsf{P}$ sends to $\textsf{V}$ the univariate polynomial
$$g_j(X_j) = \sum_{(x_{j+1}, \dots, x_{v}) \in \{0, 1\}^{v-j}}	g(r_1,\dots,r_{j-1}, X_j, x_{j+1}, \dots, x_v).$$ 
$\textsf{V}$ checks that $g_j$ is a univariate polynomial of degree at most $\deg_j(g)$,
and that $g_{j-1}(r_{j-1}) = g_j(0) + g_j(1)$, rejecting if not. 

\item $\textsf{V}$ chooses a random element $r_j \in \mathbb{F}$, and sends $r_j$ to $\textsf{P}$.

\item In round $v$, $\textsf{P}$ sends to $\textsf{V}$ the univariate polynomial
$$g_v(X_v) = g(r_1,\dots,r_{v-1},X_v).$$ $\textsf{V}$ checks that $g_v$ is a univariate polynomial of degree at most
$\deg_v(g)$, and that $g_{v-1}(r_{j-1}) = g_v(0) + g_v(1)$, rejecting if not. 

\item $\textsf{V}$ chooses a random element $r_v \in \mathbb{F}$ and evaluates $g(r_1, \dots, r_v)$ with
a single oracle query to $g$. $\textsf{V}$ checks that
$g_v(r_v) = g(r_1, \dots, r_v)$, rejecting if not.

\item If $\textsf{V}$ has not yet rejected, $\textsf{V}$ halts and accepts.
\end{itemize}
\end{varwidth}
}
\caption{Description of the sum-check protocol. $\deg_i(g)$ denotes the degree of $g$ in the $i$th variable.} 
\label{fig}
}
\end{figure}
\noindent \textbf{Properties and Costs of the Sum-check Protocol.}
The sum-check protocol satisfies perfect completeness, and has soundness error $\eps \leq \deg(g)/|\mathbb{F}|$,
where $\deg(g)$ denotes the total degree of $g$ (see \cite{lfkn} for a proof). 
There is one round of prover--verifier interaction in the sum-check protocol for each
of the $v$ variables of $g$, and the total communication is $O(\deg(g))$ field elements.

Note that as described in Figure \ref{fig}, the sum-check protocol assumes that the verifier has oracle access to $g$. However, this will not be the case in applications,
as $g$ will ultimately be a polynomial that depends on the input data stream. In order to apply the sum-check protocol in a streaming setting,
it is necessary to assume that $\textsf{V}$ can evaluate $g$ at any point $\mathbf{r}$ in small space with a single streaming pass over the input (this
assumption is made in Theorem \ref{lem::sum-check-protocol}). Alternatively,  one can have the prover
\emph{tell} the verifier $g(\mathbf{r})$, and then prove to the verifier that the value $g(\mathbf{r})$ is as claimed, using further applications of the sum-check
protocol, or heavier hammers such as the GKR protocol (As described below), which is itself based on the sum-check protocol.

\begin{theorem}[Streaming Sum Check Protocol \cite{cormode2011verifying}]
\label{lem::sum-check-protocol}
Let $g$ be a $v$-variate polynomial over $\field$, which may depend on the input stream $\tau$. Denote the degree of $g$ in variable $i$ by $\text{deg}_i(g)$. 
Assume $\textsf{V}$ can evaluate $g$ at any point $\mathbf{r} \in \field$ with a streaming pass over $\tau$, using $O(v \cdot \log |\field|)$ bits of space. 
There is an SIP for computing the function $F(\tau) = \sum_{\sigma \in \field^v} g(\sigma)$ that uses $O(v)$ messages and  $O(\sum_{i=1}^v \deg_i(g) \cdot \log|\field|)$ communication, as well as $O(v \cdot \log|\field|)$ space. 
\end{theorem}
\vspace{5mm}
For completeness, we present description of the sum-check protocol of Lund et al. \cite{lfkn} in Figure \ref{fig}.


\subsubsection{The GKR Protocol}
\label{sec:gkr}
Interactive proofs can be designed by algebrizing a circuit computing a function. 
One of the most powerful protocols of this form is due to Goldwasser et al. \cite{gkr}, and known as the GKR protocol. 
This was adapted to the streaming setting by Cormode et al. \cite{cormode2011verifying}, yielding the following result.
\begin{lemma}[\cite{gkr,cormode2011verifying}]\label{lemma:muggles} Let $\field$ be a finite field, and let $f \colon \field^{u} \rightarrow \field$ be a function of the entries of the frequency vector of a data stream (viewing
the entries as elements of $\field$).
Suppose that $f$ can be computed 
by an $O(\log(S) \cdot \log(|\field|))$-space uniform arithmetic circuit $\mathcal{C}$ (over
$\mathbb{F}$) of fan-in 2, 
size $S$, and depth $d$, with the inputs of $\mathcal{C}$ being the entries of the frequency vector. Then, assuming that $|\field| = \Omega(d \cdot \log S)$,
$f$ possesses an SIP requiring $O(d \cdot \log S)$ rounds. The total space cost is $O(\log u \cdot \log |\field|)$
and the total communication cost is $O(d \cdot \log(S) \cdot \log |\field|)$.
\end{lemma}




\section{Rectangular Matrix Multiplication and Eigenstructure}
\label{sec:verify-matr-eigenstr}

Many algorithms in data analysis require
computation of the \emph{eigenpairs} (eigenvalues and eigenvectors) of a large data
matrix. Eigenvalues of a streamed $n\times n$ matrix can be computed
approximately without a prover \cite{andoni2013eigenvalues}, but there are no
streaming algorithms to compute the \emph{eigenvectors} of a matrix because of
the output size. 

\emph{Verifying} the eigenstructure of a symmetric matrix $A$ is more difficult than merely verifying
that a claimed $(\lambda, \mathbf{v})$ is an eigenpair. This is because the prover
must convince the verifier not only that each $(\lambda_i, \mathbf{v}_i)$
satisfies $A\mathbf{v} = \lambda \mathbf{v}$, but that the collection of
eigenvectors together are orthogonal. Thus, the prover must prove that 
$V V^\top = D$ where $V$ is the collection of eigenvectors and $D$ is some diagonal
matrix. Note however that this matrix multiplication check is
\emph{rectangular}: if we wish to verify that a collection of $k$ eigenvectors
are orthogonal, we must multiply a $k \times n$ matrix $V$ by an $n\times k$ matrix $V^\top$. 

We present an annotation protocol called \IMM to verify such a \emph{rectangular} matrix
multiplication. Our protocol builds on the optimal annotations protocols for inner product and matrix-vector multiplication from \cite{chakrabarti2014annotations} and \cite{cormode2013streaming}.
We prove that our \IMM protocol obtains tradeoffs between communication and space usage that are optimal up to a factor of $\tilde{O}\left(\min\left(k, k'\right)\right)$.

\begin{theorem}
\label{thm:verify-matr-eigenstr}
Let $A$ be a $k \times n$ matrix and $B$ an $n \times k'$ matrix, both with entries in a finite field $\field$ of size $6n^3 \leq |\field| \leq 6 n^4$.
Let $(h, v)$ be any pair of positive integers such that $h \cdot v \geq n$. There is a
annotated data streaming protocol for computing the product matrix $C=A \cdot B$
with communication cost $O(k \cdot k' \cdot h \cdot \log n)$ bits and space cost $O(v \cdot \log n)$ bits.
Moreover, any (online) annotated data streaming protocol for the problem requires the product of the space and communication costs to be at least $\Omega\left( (k + k') \cdot n\right)$.
\end{theorem}



\begin{proof}
To present the upper bound, we first recall the inner product protocol of Chakrabarti et al. \cite{chakrabarti2014annotations}. Given input vectors $a, b \in \field^n$, the verifier 
in this protocol treats the $n$ entries of $a$ and $b$ as a grid $[h] \times [v]$, and
considers the unique bivariate polynomials $\widetilde{a}(X, Y)$ and $\widetilde{b}(X, Y)$ over $\mathbb{F}$ of degree at most $h$ in $X$ and $v$ in $Y$ satisfying
$\widetilde{a}(x, y) = a(x, y)$ and $\widetilde{b}(x, y) = b(x, y)$ for all $(x, y) \in [h] \times [v]$.  
The verifier picks a random $r \in \mathbb{F}$, and evaluates $\tilde{a}(r, y)$ and $\tilde{b}(r, y)$ for all $y \in [v]$. As observed in \cite{chakrabarti2014annotations},
the verifier can compute $\tilde{a}(r, y)$ for any $y \in [v]$ in space $O(\log |\field|)$, with a single streaming pass over the input. Hence, the verifier's total
space usage is $O(v \cdot \log |\field|)$. 
The prover then sends a univariate polynomial
$s(X)$ of degree at most $h$, claimed to equal $g(X) = \sum_{y\in [v]} \tilde{a}(X, y) \cdot  \tilde{b}(X, y)$. The verifier accepts $\sum_{x \in [h]} s(X)$ as
the correct answer if and only if $s(r) = \sum_{y \in [v]} \tilde{a}(r, y) \cdot \tilde{b}(r, y)$. 

Returning the matrix multiplication, let us denote the rows of $A$ by $\mathbf{a}_1, \dots, \mathbf{a}_k$ and the columns of $B$ by $\mathbf{b}_1, \dots, \mathbf{b}_k$. Notice
 that each entry $C_{ij}$ of $C$ is the inner product of $\mathbf{a}_i$ and $\mathbf{b}_j$. 

\medskip
\noindent \textbf{The prover's computation.} In our matrix multiplication protocol, the prover simply runs the above inner product protocol $k \cdot k'$
times, one for each entry $C_{ij}$ of $C$. This requires sending $k \cdot k'$ polynomials, $s_{ij}(X) \colon (i, j) \in [k] \times [k']$, each of degree at most $h$. Hence, the total communication
cost is $O(k \cdot k' \cdot h \cdot \log n)$.

\medskip
\noindent \textbf{The verifier's computation while observing entries of $A$.}
The verifier picks a random $\alpha$ and computes, for each $y \in [v]$, the quantity $s_y := \sum_i \tilde{a}_i(r, y) \alpha^i$. 
Using standard techniques \cite{chakrabarti2014annotations}, the verifier can compute each $s_y$ with a single streaming pass over the entries of $A$, in $O(\log n)$ space. Hence, the verifier
can compute all of the $s_y$ values in total space $O(v \cdot \log n)$.

\medskip
\noindent \textbf{The verifier's computation while observing entries of $B$.} 
For each $y \in [v]$, the 
verifier computes the quantity $s'_y := \sum_{j\in k'} \tilde{b}_j(r, y) \alpha^{k \cdot j}$. 
The reason that we define $s'_y$ in this way is because it ensures that $s_y \cdot s'_y =\sum_{ (i,j) \in [k] \times [k']} \tilde{a}_i(r, y) \cdot \tilde{b}_j(r, y) \alpha^{k\cdot j + i}$, which is just a fingerprint of the set of values $\{\tilde{a}_i(r, y) \cdot \tilde{b}_j(r, y)\}$ as $(i, j)$ ranges over $[k] \times [k']$.

To check that all $s_{ij}$ polynomials are as claimed, the verifier does the following. As the verifier reads the $s_{ij}$ 
polynomials, she computes a fingerprint of the $s_{i, j}(r)$ values, i.e., the verifier computes $\sum_{i, j} s_{i, j}(r) \cdot \alpha^{j \cdot k +i}$. The verifier 
checks whether this equals $\sum_y (s_y \cdot s'_y)$. If so, the verifier is convinced that $A_{ij} = \sum_{x \in [h]} s_{ij}(x)$ for all $(i, j) \in [k] \times [k']$. 
If not, the verifier rejects. 

\medskip
 \noindent \textbf{Proof of completeness.}
If the $s_{i, j}$ polynomials are as claimed, then:
\begin{align*} \sum_{i,j\in [k] \times [k']} g_{i, j}(r) \cdot \alpha^{j \cdot k +i} = \sum_{i,j\in [k] \times [k']}  \sum_{y\in[v]} \tilde{a}_i(r, y) \cdot  \tilde{b}_j(r, y) \alpha^{j\cdot k +i}\\
 = \sum_{y\in[v]} \sum_{i,j\in[k] \times [k']}  \tilde{a}_i(r, y) \cdot  \tilde{b}_j(r, y) \alpha^{j\cdot k +i}  = \sum_{y\in [v]} s_y \cdot  s'_y.\end{align*}
\medskip \noindent \textbf{Proof of soundness.}
If any of the $s_{i, j}$ polynomials are \emph{not} as claimed (i.e., if $s_{ij}(X) \neq g_{ij}(X)$ as formal polynomials), 
then with probability at least $1-h/|\mathbb{F}|$ over the random choice of $r \in \mathbb{F}$, it will hold that $s_{i, j}(r) \neq g_{ij}(r)$. 
In this event the verifier will wind up comparing the fingerprints of two different vectors, namely the $k \cdot k'$-dimensional vector whose $(i, j)$'th entry is $s_{i, j}(r)$, and the $k \cdot k'$-dimensional vector vector whose $(i, j)$'th entry is $\sum_{y \in [v]} \tilde{a}_i(r, y) \cdot \tilde{b}_j(r, y).$ These fingerprints will disagree with probability at least $1-k \cdot k'/|\mathbb{F}|$. Hence,
the probability that the prover convinces the verifier to accept is at most $h/|\mathbb{F}| + k \cdot k'/|\mathbb{F}|$. If $|\mathbb{F}| \geq 100 \cdot h \cdot k \cdot k'$, the soundness error will be bounded
by $1/50$.

\medskip \noindent \textbf{Lower bound.} 
Cormode et al. \cite{cormode2013streaming} proved a lower bound on the cost of (online) annotated data streaming protocols for \emph{matrix-vector} multiplication (i.e., for multiplying a $k \times n$ matrix $A$ by an $n \times 1$ matrix $B$). Specifically, their argument implies that if $A$ is $k \times n$, then any protocol for multiplying $A$ by a vector must have the product of the space and communication costs be at least $\Omega(k \cdot n)$. The claimed lower bound follows if $k > k'$ (the case of $k < k'$ is analogous).

\end{proof}
\noindent \textbf{On \textsf{V}'s and \textsf{P}'s runtimes.} 
Using Fast Fourier Transform techniques (cf. \cite[Section 2]{cormode2012practical}), the prover in the protocol of Theorem \ref{thm:verify-matr-eigenstr} can run in $O(k \cdot k' \cdot n \log n)$ total time, assuming the total number of updates to the input matrices $A$, $B$ is $O(k \cdot k' \cdot n \log n)$. The verifier can run in time $O(\log n)$ per stream update.

\vspace{5mm}
\noindent \textbf{The Eigenpair Verification Protocol.}
\label{sec:eigen}
\label{sec:protocol-1}
We now show how to use Theorem \ref{thm:verify-matr-eigenstr} to verify that a claimed set of $k$
eigenvalues and eigenvectors are indeed (approximate) eigenpairs of a given symmetric integer input matrix $A$. The protocol is cleanest to present assuming the entries of all of the claimed eigenvectors are integers, in which case the protocol can verify that the vectors are \emph{exact} eigenvectors. We explain how to handle the general case at the end of the section.

\vspace{2mm}
\noindent \textbf{The case where all claimed eigenvectors have integer entries.} The eigenpair verification protocol invokes \IMM twice. In the first invocation, \IMM is used to simultaneously verify
that all claimed eigenpairs are indeed eigenpairs. Specifically, the \IMM protocol is used to compute $C=A \cdot V$, where $V$ is the matrix
whose $i$th column equals the $i$th claimed eigenvector $\mathbf{v}_i$. The verifier use fingerprints to check that $C=V \cdot D$,
where $D$ is the diagonal matrix with entries corresponding to the claimed eigenvalues. In the second invocation, \IMM is used to check that the claimed eigenvectors are orthogonal, by verifying 
that $V^\top  V = D'$ for some diagonal matrix $D'$ provided by the prover.
Note that in both invocations of the \IMM protocol, the verifier does not have the space to explicitly store the matrix $V$. Fortunately, storing $V$ is not necessary, as within both invocations of the \IMM protocol, $V$ is treated as part of the input stream, and the \IMM protocol does not require the verifier to store the input. 
%
%
\vspace{2mm}

\noindent \textbf{The general case.} 
We now sketch at a high level how to handle the general case, in which the entries of the claimed eigenvalues are not integers (note that since $A$ is symmetric, the entries of all of its eigenvalues can be taken to be real). The protocol guarantees in this general case that, for any desired error parameter $\eps$, each claimed eigenpair $(\lambda_i, \mathbf{v}_i)$ satisfies
$\|A \mathbf{v}_i - \lambda_i \mathbf{v}_i\|_2 \leq \eps$.
The approach we take to handle non-integer entries is exactly as in the eigenvalue-verification protocol of Cormode et al. \cite{cormode2013streaming}. Specifically, we reduce to the integer case by requiring the prover to round the entries of all claimed eigenvectors and eigenvalues to an integer multiple of $\eps'$ for some sufficiently small value $\eps'$, in such a way that the resulting eigenvectors are exactly orthogonal. It is straightforward to show that 
there is some $\eps' = 1/\poly(n, \eps^{-1})$ such that the rounding changes each entry of $A \mathbf{v}_i$
by at most $\eps/n^2$.
This ensures that
the matrix $V/\eps'$ is has integer entries, all bounded in absolute value by $\poly(n/\eps)$. Hence,
each entry of $V/\eps'$ can be identified with an element of a finite field of size $\poly(n, \eps^{-1})$,
and we can apply the integer matrix multiplication protocol to compute $A \cdot (V/\eps')$ and $(V/\eps')^\top  (V/\eps')$. The verifier checks that the latter result is a diagonal matrix, guaranteeing that the claimed eigenvectors are orthogonal. Given the former result, it is straightforward for the prover to convince the verifier that each entry of the former matrix is close enough to $(V/\eps') \cdot D$ to ensure that $\|A \mathbf{v}_i - \lambda_i \mathbf{v}_i\|_2 \leq \eps$.
\begin{theorem} \label{eigentheorem}
Let $A$ be a symmetric $n\times n$ integer matrix with entries bounded in absolute value by $\poly(n)$.  Let $k$ be an
integer, let $h$ and $v$ be positive integers satisfying $h \cdot v \geq n$ and let $\eps>0$ be an error parameter. Then there is an annotated data streaming protocol for verifying that a
collection of $k$ eigenpairs $(\lambda_i, \mathbf{v}_i)$ are orthogonal, and each satisfies
$\|A \mathbf{v}_i - \lambda_i \mathbf{v}_i\|_2 \leq \eps$. The total communication cost is $O(k^2 \cdot h \cdot \log(n/\eps))$
and the verifier's space cost is $O(v \cdot \log(n/\eps))$.
\end{theorem}




\section{Shape Analysis in a Few Rounds}
\label{sec:verify-optim-few}
\label{sec:constant}
In this section, we give 3-message SIPs of polylogarithmic cost for finding an \ac{meb} and computing the width of a point set. The key here is to identify a sparse dual witness that proves optimality (or near-optimality) of the claimed (primal) solution and then check feasibility of both primal and dual solutions. 
We show how the verifier can perform both feasibility checks via a careful reduction to an instance of the \range\ problem.
\subsection{Verifying Minimum Enclosing Balls}
\label{sec:minim-encl-balls}
\label{sec:meb}

Consider the Euclidean $k$-center problem with $k=1$, otherwise known as the \ac{meb}: given a set of $n$ points $P \subset \mathcal{U}$ in which $\mathcal{U} = {[m]}^d$, find a ball $B^*$ of minimum radius that encloses all of them. 

The \ac{meb} presents an interesting contrast between our model and the classical streaming model. It is known that \emph{no} streaming algorithm that uses poly$(d)$ space can approximate the \ac{meb} of a set of points to better than a factor of $\sqrt{2}$ by a coreset-based construction and $\frac{1+ \sqrt{2}}{2}$ in general \cite{agarwal2010streaming}. Also, the best streaming \emph{multiplicative} $(1+\epsilon)$-approximation for the \ac{meb} uses $O({(1/\epsilon)}^{\frac{d}{2}})$ space \cite{chan06}. 

\subsubsection{The Protocol}
The prover reads the input and sends the (claimed) minimum enclosing ball $B$. Our protocol reduces checking feasibility and optimality of $B$ to carefully constructed instances of the \range problem.

\vspace{-4mm}
\paragraph{Checking Feasibility.}
We consider a new range space, in which the range set $\mathcal{B}$ is defined to consist of all balls with radius $j  \colon j \in \{0, 1, \dots, m^d\}$ and with centers in $[m]^d$. Notice that $|\mathcal{B}|= O(m^{2d})$.
Using the protocol for \range (Theorem \ref{lemma:index}), we can verify that the claimed solution $B$ does in fact cover all points (because this will hold if and only if the range count of $B$ equals the cardinality of the input point set $|P| = n$). 
\vspace{-3mm}
\paragraph{Checking Optimality.}
We will make use of the following well known fact about minimal enclosing balls, which was used as the main idea for developing an approximation algorithm for furthest neighbour problem, by Goel et al. \cite{DBLP:conf/soda/GoelIV01}:
\vspace{-1mm}
\begin{lemma}{\label{meb}}
  Let $B^*$ be the minimal enclosing ball of a set of points $P$ in $\mathbb{R}^d$. Then there exist at most $d+2$ points of $P$ that lie on the boundary $\partial B^*$ of $B^*$ and contain the center of $B^*$ in their convex hull. 
\end{lemma}
\paragraph{Putting it all Together.}
The complete 3-message \ac{meb} protocol works as follows.

\vspace{-2mm}
\begin{enumerate}
\item \textsf{V} processes the data stream for \range (with respect to $\mathcal{B}$ and $P$). 
\item \textsf{P} computes the \ac{meb} $B^*$ of $P$, then rounds the center $c$ of the $\ac{meb}$ to the nearest grid vertex. Denote this vertex by $c^*$. \textsf{P} sends $c^*$ to $\textsf{V}$, as well as the radius $r$ of $B^*$, and a subset of points $T \in P$ in which $\ac{meb}(T) = \ac{meb}(P)$. (Note that based on Lemma \ref{meb}, $|T| \leq d+2$ suffices).
\item \textsf{V} first computes the center $c$ of the \ac{meb} for the subset $T$ and checks if $c^*$ is actually the rounded value of $c$. Then $\textsf{V}$ runs a \range protocol with \textsf{P} to verify that the ball of radius $r+ 1$ and center $c^*$ contains all of the input points. It then runs multiple copies of \pq to verify that the subset $|T| \leq d+2$ points provided by \textsf{P} are actually in the input set $P$. 
\end{enumerate}
 \vspace{-3mm}
\begin{theorem} \label{mebtheorem}
There exists a 3-message SIP for the Minimum Enclosing Ball (\ac{meb}) problem with communication and space cost bounded by $O(d^2 \cdot \log^2 m) $. \end{theorem}
\noindent \textbf{On \textsf{V}'s and \textsf{P}'s runtimes.} Assuming the distance function $D$ under which the instance of $\ac{meb}$ is defined satisfies mild 
``efficient-computability'' properties, both $\textsf{V}$ and $\textsf{P}$ can be made to run in total time $\text{polylog}(m^d)$ per stream update in the protocol of Theorem \ref{mebtheorem}. Specifically, it is enough that 
for any point $\bx \in P$, there is a De-Morgan formula of size $\text{polylog}(m^d)$ that takes as input the binary representation of
a ball $B \in \mathcal{B}$ and outputs $1$ if and only if $\bx \in B$. 
Under the same assumption on $D$, the prover $\textsf{P}$ can be made to run in time $T + n \cdot \text{polylog}(m^d)$, where $T$ is the time required to find the MEB of the input point set $P$. 
For details, see the full description of the $\pq$ protocol of \cite{chakrabartiinteractivity}.

\subsubsection{Streaming lower bounds on the grid} 
We note that restricting the points to a grid does not make the \ac{meb} problem easier for a streaming algorithm. 
Here we show that lower bound for streaming \ac{meb} due to Agarwal and Sharathkumar \cite{agarwal2010streaming} can be modified to work even if the points lie on a grid. 
The key lemma in Agarwal and Sharathkumar's lower bound is a construction of a collection of almost orthogonal vectors that are centrally symmetric. Let $S^{d-1}$ denote the unit sphere in $\reals^d$.



\begin{lemma}[Agarwal and Sharathkumar \cite{agarwal2010streaming}]
There is a centrally symmetric point set $K \subset S^{d-1}$ of size $\Omega(\exp(d^{\frac{1}{3}}))$ such that for any pair of distinct points $p, q \in K$ if $p \neq -q$, then
\begin{align}
\sqrt{2}(1 - \frac{2}{d^{\frac{1}{3}}}) \leq \|p-q\| \leq \sqrt{2}(1 + \frac{2}{d^{\frac{1}{3}}})
\end{align}
\end{lemma}
This point set is then used by an adversary to ``defeat'' any algorithm claiming a $\sqrt{2}-\delta$ approximation. Note that the ``almost orthogonal'' property follows from the observation that for unit vectors $p,q$, $\|p-q\|^2 = 2 - 2 \langle p, q\rangle$ and therefore the condition of the lemma above implies that $\langle p,q \rangle \le \frac{4}{d^{\frac{1}{3}}}$

It turns out that this ``almost-orthogonal'' property can be achieved by vectors with integer coordinates. The proof is in the same spirit of the proofs that sign matrices can be used in the Johnson-Lindenstrauss lemma, and follows from an observation by Ryan O'Donnell \cite{mathoverflow}. We recreate the proof here for completeness. 

\begin{lemma}[Bernstein's inequality]
Let $X_1, \ldots, X_d$ be independent Bernoulli variables taking values in $\{+1, -1\}$ with equal probability. Then 
\[ \text{Pr}[|\frac{1}{d}\sum_i X_i| \ge \epsilon] \le 2\exp\left(-d\epsilon^2/\left(2\left(1+\epsilon/3\right)\right)\right). \]  
\end{lemma}
\begin{lemma}\label{innerprod} Let $t=\exp(\frac{\epsilon^2 d}{4})$ . Let $\mathbf{u}_1, \dots, \mathbf{u}_t$ be random vectors in which each entry is set to $1/\sqrt{d}$ or $-1/\sqrt{d}$, with probability $\frac12$ each.
 There is a positive probability of $|\langle\v{u_i}, \v{u_j}\rangle| \le \epsilon$ holding for all $i \neq j$. 
\end{lemma}
\begin{proof}

We define variables $x_{ij}$ as the Bernoulli variables corresponding to Lemma \ref{innerprod}, where $i \le k, j \le d$. That is, define the $x_{ij}$ variables such that:
\[ \mathbf{u}_i = \frac{1}{\sqrt{d}}(x_{i1}, \ldots, x_{id}).\]

We want to analyze the behavior of $\langle \v{\mathbf{u}}_i, \v{\mathbf{u}}_j \rangle$. 
Set $Y^{ij}_k = x_{ik}x_{jk}$ and write $\langle \v{\mathbf{u}}_i, \v{\mathbf{u}}_j \rangle$ as $\frac{1}{d}\sum_{k} Y^{ij}_k$. Note that for each $i$, $j$, and $k$, $Y^{ij}_k$ is a Bernoulli variable with range $\{-1, +1\}$, and for any fixed $i,j$, the variables $Y^{ij}_k$ are independent. 
Therefore, we can apply Bernstein's inequality to the collection $\{Y^{ij}_k\}$ for a fixed $i,j$. 

For simplicity, assume that $\epsilon \le 1$. Then Bernstein's inequality implies that
\[ \text{Pr}[|\langle \v{u}_i, \v{u}_j\rangle|\ge \epsilon] \le 2\exp(-d\epsilon^2/4). \]
It follows that the probability that $|\langle \v{\mathbf{u}}_i, \v{\mathbf{u}}_j\rangle| \ge \epsilon$ is at most $2\exp(-d\epsilon^2/4) $. Now if we set $t = \exp(\frac{d\eps^2}{4})$, 
then this probability value equals $\frac{2}{t^2}$ by choice of $t$ and hence by taking a union bound over at most ${t \choose 2} \le \frac{t^2}{2}$ pairs of $(i, j)$ we conclude that there is a positive probability of $|\langle \v{\mathbf{u}}_i, \v{u}_j\rangle]\le \epsilon$ holding for all $i \neq j$.
\end{proof}



\subsection{Verifying the Width of a Point Set}

\label{sec:protocol-one-slab}
Let the width of a point set be the minimum distance between two parallel
hyperplanes that enclose it. 
Like the \ac{meb} problem, the width of a point set can be approximated by a streaming algorithm
using $O(1/\epsilon^{O(d)})$ space \cite{chan06}, without access to a prover. 

We present a similar protocol for verifying the width of a point set as follows:
We describe an efficient constant-round SIP to \emph{exactly compute}
the width of a point set. As before, we study the problem in the discrete setting, i.e., we assume that the data stream elements are a subset of points over a grid structure $\mathcal{U}= {[m]}^d$. 
Let $\mathcal{R}$ denote the set of all the ranges defined by single slab (i.e., each range consists of the area between some two parallel hyperplanes).

\subsubsection{Certificate of Optimality}
Given a slab $S$ that is claimed to be a minimal-width slab covering the input point set $P$, the following lemma (akin to Lemma~\ref{meb}) guarantees the existence of a
sparse witness of optimality for $S$. 

\begin{lemma} \label{points-cover}
Given the input point set $P$ in $d$-dimensions, every optimal-width single slab $S$ consisting of the area between parallel hyperplanes $h_1, h_2$ covering $P$ can be described by a set of $k+k'=d+1$ points from input point set $P$, in which $k$ points lie on the hyperplane $h_1$ and $k'$ points lie on the hyperplane $h_2$.
\end{lemma}

\begin{proof}
We express $S$ as an optimal solution to a certain linear program. We then infer the existence of the claimed witness of optimality for $S$
via strong linear programming duality and complementary slackness.

Assume the two hyperplanes specifying $S$ are of the form $h_1: \langle \bw, \bx \rangle = 1$ and $h_2:  \langle \bw, \bx \rangle = \ell$, where $\bw \in \mathbb{R}^d$. Then the pair $(\bw, \ell)$ 
corresponds to an optimal solution of the following linear program: 
\begin{align*}
  \min & \quad \ell \\
\text{s.t. } \forall i \in \{1, \dots, |P|\}& \quad \langle \bw, \bx_i\rangle \geq 1 \\
\forall i \in \{1, \dots, |P|\} & \quad \langle \bw, \bx_i\rangle \le \ell
\end{align*}

We write the LP in the standard form:
\begin{align*}
  \max & \quad -\ell \\
\text{s.t. } \forall i \in \{1, \dots, |P|\} & \quad (-\bx_i^T) \cdot \bw \geq 1 \\
\forall i \in \{1, \dots, |P|\} & \quad \bx_i^T \cdot \bw -\ell \le 0
\end{align*}


Let $x_{ij}$ denote the $j$th entry of input point $\bx_i \in [m]^d$. Standard manipulations reveal the dual.
\begin{align*}
  \min & \quad \sum_{i=1}^{|P|} y_i  \\
\text{s.t. } \forall j \in \{1, \dots, d\} & \quad \sum_{i=1}^{|P|} (y_i - z_i) x_{ij} = 0. \\
& \quad \sum_{i=1}^{|P|} z_i = 1.
\end{align*}

Let $\by=(y_1, \dots, y_{|P|})$ and $\bz = (z_1, \dots, z_{|P|})$ denote an optimal solution to the above dual.

\paragraph*{Claim 1.} For any $i$, $y_i$ and $z_i$ cannot both be nonzero.

\begin{proof}
By complementary slackness, $y_i$ and $z_i$ are both nonzero only if both of the corresponding primal inequalities are tight, which can only hold if the width is zero. 
\end{proof}

\paragraph*{Claim 2.} In total, the number of nonzero entries in $y$ and $z$ must be at least $d+1$.  

\begin{proof}
Fix any $j \in \{1, \dots, d\}$ and consider the constraint \begin{equation} \sum_{i=1}^{|P|} y_i x_{ij} = \sum_{i=1}^{|P|} z_i x_{ij} \label{ugh} \end{equation}
 from the dual. Note that by Claim 1, all the $x_{ij}$'s with a non-zero coefficient $y_i$ in the left hand side of Equation \eqref{ugh}
are distinct from the $x_{ij}$'s with a non-zero coefficient $z_i$ on the right hand side of Equation \eqref{ugh}. Suppose by way of contradiction that there are at most $d$ nonzero entries in total in $y$ and $z$. 
Fix one such non-zero entry, say, $z_k$. We can rewrite Equation \eqref{ugh} as:
\begin{align*}
z_k x_{kj} = \sum_{i=1}^{|P|} y_i x_{ij} - \sum_{i \neq k} z_i x_{ij}
\end{align*}
and by dividing by $z_k$ and relabeling the coefficients, we get:
\begin{align*}
x_{kj} = \sum_{i=1}^{|P|} \alpha_i x_{ij}
\end{align*}
for some coefficients $\alpha_1, \dots, \alpha_{|P|} \in \mathbb{R}$,
where at most $d-1$ of the $\alpha_i$'s are non-zero. But this says that there exist $d$ points not in general position, which is a contradiction. Therefore Claim 2 is true.
\end{proof}
\end{proof}

Now using Lemma \ref{points-cover}, we can give the following upper bound for the size of the range set $\mathcal{R}$, in the one-slab problem on $\mathcal{U} = {[m]}^d$.
\begin{lemma}\label{rangesize}
Given a grid $\mathcal{U}= {[m]}^d$, the size of the range set $\mathcal{R}$ consisting of all slabs is $O(m^{d^2+d})$.
\end{lemma}

\begin{proof}
Based on Lemma \ref{points-cover}, each slab on the grid ${[m]}^d$ can be determined by two parallel hyperplanes including $d+1$ points. Thus we have:
\begin{align*}
|\mathcal{R}| = \sum_{k=1}^{d+1} {|\mathcal{U}| \choose k} {|\mathcal{U}| \choose d+1-k}
&\leq \sum_{k=1}^{d+1} {m^d \choose k} {m^d \choose d+1-k}\\
&= {2m^d \choose d+1} = O(m^{d^2 + d})
\end{align*}
\end{proof}

\subsubsection{The Protocol}
The protocol works as follows:
\begin{enumerate}
\item \textsf{V} processes the data stream as if for a \range query with respect to $\mathcal{R}$, defined as the set of single slabs including $k+k' = d+1$ points. 
\item \textsf{P} returns a candidate slab $S$ consisting of two parallel hyperplanes $h_1, h_2$, claimed as the slab with minimum width which covers all the points $P$ in input data stream.
\textsf{P} also sends a set $T_1$ of $k$ points and a set $T_2$ of $k'$ points claimed to satisfy the properties of Lemma \ref{points-cover}.
\item \textsf{V} verifies that if $k+k' = d+1$, checks that all points in $T_1$ lie on $h_1$ and all points in $T_2$ lie on $h_2$, and runs the \pq protocol $d+1$ times to check 
that all points in $T_1 \cup T_2$ actually appeared in the input set $P$.
\item \textsf{V} initiates a \range query for the range corresponding to the slab $S$, and verifies that the answer is $n = |P|$, i.e., that $S$ covers all the input points. 
\end{enumerate}

Perfect completeness of the protocol is immediate from Lemma \ref{points-cover} and the completeness of the \pq and \range protocols. 
The soundness error of the protocol is at most $(d+2) \cdot \eps_s$, where $\eps_s \leq \frac{1}{3(d+2)}$ is an upper bound on the soundness errors of the \pq and \range protocols.
To see this, note that if $T_1$ and $T_2$ are as claimed, then there is no slab of width less than that of $S$ covering the input points. And the probability
that the verifier accepts when $T_1$ and $T_2$ are not as claimed is bounded by $(d+2) \cdot \eps_s$, via a union bound over all $(d+1)$ invocations of the \pq protocol and the single invocation of the \range protocol.
Theorem~\ref{thm:width} follows.

In the protocol of Theorem \ref{thm:width}, the prover and verifier can be made to satisfy the same runtimes bounds as in the $\ac{meb}$ protocol of Section \ref{sec:meb},
assuming the distinct function $D$ satisfies the same ``efficient computability'' condition discussed there.  

\begin{theorem}
\label{thm:width}
Given a stream of $n$ input points from $\mathcal{U}= {[m]}^d$, there is a three-message SIP for verifying the width of the input with space and communication cost bounded by $O(d^4 \cdot \log^2 m)$. 
\end{theorem}







\subsection{Verifying Approximate Metric $k$-Centers}
\label{sec:verify-appr-metr}
Using the same ideas as for the \ac{meb}, we can
verify a 2-\emph{approximation} to the metric $k$-center problem. At a high
level, the SIP verifies the correctness of the witness
produced by running Gonzalez' approximation algorithm for metric $k$-center
clustering \cite{gonzalez1985clustering}: 
namely, $k+1$
 points that are at least distance $r$ apart (where $r$ is the claimed
 2-approximate radius). This sparse witness can be verified using the \pq
 protocol. 
 
 Here we describe the formalization of the metric $k$-Center problem and the protocol in details and then the main result follows.
 
 \noindent
A $k$-center clustering of a set of points $p_1, \ldots, p_n$ in a metric space $(X, d)$ is a set of $k$ centers $\mathcal{C} = \{c_1, \ldots c_k\}$. The cost of such a clustering is 
\begin{align*}
\text{ cost}(C) = \max_i \min_j d(p_i, c_j). 
\end{align*}

\begin{definition}
Let $(X, d)$ be a metric space. Let $p_1, p_2, \ldots, p_n, k$ be a stream of points from $(X, d)$ followed by parameter $k$. 
An SIP computing a $2$-approximation for the metric $k$-center problem with completeness error $\eps_c$ and soundness error $\eps_s$ has the following form.
The prover begins the SIP by claiming that there exists a $k$-center clustering of cost $r^*$. 
\begin{itemize}
\item If this claim is true, the verifier must accept with probability at least $1-\eps_c$.
\item If there is no $k$-center clustering of cost at most $r^*/2$, the verifier must reject with probability at least $1-\eps_s$. 
\end{itemize}
\end{definition}

It is easy to provide a protocol that works deterministically if the verifier is not required to process the input in a streaming manner. This is the standard 2-approximation algorithm of Gonzalez: the prover provides
\begin{description}
\item[Proof of Feasibility.] A set of \emph{centers} $c_1, \ldots, c_k$ satisfying $\max_i \min_j d(p_i, c_j) \le r^*$ and
\item[Proof of Approximate Optimality.] A set of $k+1$ points $u_1, u_2, \ldots, u_{k+1}$ from the stream with the promise that $\min_{i,j} d(u_i, u_j) \geq r^*$.
\end{description}
This guarantees a $2$-approximation by the standard argument relying on the triangle inequality \cite{gonzalez1985clustering}. The verifier can easily check that the relevant conditions hold. 

\paragraph*{The SIP.}
\label{sec:protocol}

Let $B_{d, r}(x) = \{ y \in X \mid d(x, y) \le r\}$ denote a ball of radius $r$ with center $x$ in the metric space. We define a range space $\mathcal{R}$ consisting of all unions of $k$ balls of radius $r$ for all values of $r$: 
\[ \mathcal{R} = \{ \cup_{z \in Z} B_{d, r}(z) \mid Z \subset X, |Z| = k, \exists x,y \in X, d(x,y) = r\} \]
Note that $|\mathcal{R}| = O(m^{k+2})$, where $m$ is the size of metric space, i.e. $|X| = m$.

The protocol works as follows:
\begin{enumerate}
\item \textsf{V} processes the data stream as if for a \range query using range space $\mathcal{R}$, as well as for $k+1$ parallel \pq queries. 
\item \textsf{P} returns a candidate clustering $c_1, c_2, \ldots, c_k$ with the claimed cost $r^*$, as well as $k+1$ points $u_1, \dots, u_{k+1}$ from the stream witnessing (approximate) optimality. 
\item \textsf{V} initiates a \range query for the range $\cup_{i=1}^k B_{d, r^*}(c_i)$ and verifies that the answer is $n=|P|$. 
\item \textsf{V} verifies that the distance between all distinct pairs of points $(u_i, u_j)$ is at least $r^*$, and invokes $(k+1)$ \pq queries to ensure that each $u_i$ appeared in the input stream.
\end{enumerate}

The correctness of the protocol follows from the correctness of Gonzalez's algorithm and Theorem~\ref{lemma:index}. Note that approximating metric $k$-center to within a factor of $2-\epsilon$ is NP-hard \cite{feder1988optimal}. The above protocol is a streaming variant of an \textsf{MA} protocol. Under the widely-believed assumption that $\textsf{MA}=\textsf{NP}$, there is no $2-\epsilon$ approximation for metric $k$-center with a polynomial-time verifier,
regardless of whether the verifier processes the input in a streaming manner.

As with our protocols for the \ac{meb} problem and computing the width of a point set, \textsf{V} and \textsf{P} can be made to run in quasilinear time if the metric $d$ satisfies mild efficient-computability properties. 


\begin{theorem}\label{metric k-center}
Let $(X, d)$ be a metric space in which $|X| =m$. Given an input point set $|P|=n$ from $(X, d)$, there is a streaming interactive protocol for verifying $k$-center clustering on $P$ with space and communication costs bounded by $O(k+ \log(|\mathcal{R}|) \cdot \log (n \cdot|\mathcal{R}| )) $, in which $|\mathcal{R}| \leq m^{k+2}$.
\end{theorem}



\section{SIPs for General Clustering Problems}
\label{sec:verify-clust-probl}
In this section, we give SIPs for two very general clustering problems: the $k$-center problem, and the $k$-slab problem.
In the $k$-center problem, given a set of $n$ points in $[m]^d$, the goal is to find $k$ centers so as to minimize the maximum point-center distance. 
In the $k$-slab problem, the goal is instead to find $k$ \emph{hyperplanes} so as to minimize the maximum point-hyperplane distance.
\subsection{$k$-Slabs}
\label{sec:k-slabs}
\label{sec:kslab}
We first consider the $k$-slab problem. Even when $k=2$ (and $d =3$), this problem appears to be difficult to solve efficiently without access to a prover: in fact, it was shown that this problem does \emph{not} admit a core set for arbitrary inputs \cite{nocore}. Later, Edwards et al. \cite{edwards2005no} showed that if the input points are from $\mathcal{U} = {[m]}^d$ (as in our case), then there exists a coreset with size at most ${(\frac{\log m}{\epsilon})}^{f(d, k)}$ (exponential in dimension $d$), which provides a $(1+\epsilon)$-approximation to $k$-slab problem. However, $k$-slab problem does \emph{not} admit a streaming algorithm to the best of our knowledge. 
As before, we can think of a ``cluster'' as described not by a single hyperplane, but as the region between two parallel hyperplanes that contain all the points in that cluster. The \emph{width} of the cluster is the distance between the two hyperplanes. We now think of the $k$-slab objective as minimizing the maximum width of a cluster, a quantity we call the \emph{width} of the $k$-slab. 
\paragraph{Defining the Relevant Range Space.}
Each slab can be described by $d+1$ points (that define the hyperplane) in $\mathcal{U} = {[m]}^d$ and a width parameter. A $k$-slab is a collection of $k$ of such slabs. Let $\Re$ be the range space consisting set of all $k$-slabs. This range space has size  $|\Re| = m^{kd^2 + 2kd}$. For any $k$-slab $\sigma \in \Re$, let $w(\sigma)$ denote its width. 
We will assume a canonical ordering of the ranges $\sigma_1, \sigma_2, \ldots,$ in increasing order of width (with an arbitrary ordering among ranges having the same width), as well as an effective enumeration procedure that given an index $i$ returns the $i^{\text{th}}$ range in the canonical order. We will also assume the existence of a mapping function $\mathcal{M}: \reals \to \{-1, \dots, |\Re|-1\}$ which maps a width value $w$ to the smallest index $i$ such that $w(\sigma_i) = w$, and to the null value $-1$ otherwise. Notice that the verifier can compute this mapping function by explicit enumeration, using only enough space to store one range. 
\paragraph{Stream Observation Phase of the SIP.}
Let $\tau = (p_1, p_2, \ldots, p_n) $ be the stream of input points. As the verifier sees the data points, it generates a \emph{derived stream} $\tau'$ as follows.
For each point $p_i$ in the actual input stream $\tau$, $\textsf{V}$ inserts into $\tau'$ all $k$-slabs $\sigma \in \Re$ which contain the point $p_i$. Notice
that $\tau'$ is a deterministic function of $\tau$, and hence the prover $\textsf{P}$, who sees $\tau$,
can also materialize $\tau'$, with no communication from \textsf{V} to \textsf{P} required to specify $\tau'$.\footnote{The running time cost increase for the mapping function and the derived stream can be avoided by observing that the frequency vector $f_\mathbf{a}$ is not arbitrary, since it tracks membership in ranges. This trick is described in \cite{chakrabartiinteractivity} and allows us to modify the extension polynomial used to report entries of the vector without needing to write down the explicit derived stream. Also see the discussion in Section~\ref{sec:verify-optim-few}.}
Note that the frequency $f_\sigma$ of the range $\sigma$ in this derived stream $\tau'$ is the number of points that $\sigma$ contains. 
\paragraph{Proving Feasibility.}
\label{sec:feasibility}

After the stream $\tau$ has passed, $\textsf{P}$ supplies a candidate $k$-slab $\sigma^*$ and claims that this has optimal width $w^* = w(\sigma^*)$. By applying the \range protocol from Theorem \ref{lemma:index} 
to the derived stream $\tau'$, \textsf{V} can check that $f_\sigma^* = n$ and is therefore feasible. This feasibility check requires only 3 messages. 
\paragraph{Optimality.} Proving optimality is more involved and for that we use GKR protocol as follows.

\label{sec:optimality}
\label{app:optimal}
The verifier must check if the optimal width is $w$ as claimed by the prover. Given a subset $S \subseteq \Re$ of $k$-slabs, let 
$\ind_S \colon \{0, 1\}^{\log |\Re|} \rightarrow \{0, 1\}$ denote the indicator function that evaluates to $1$ on the binary representation of a range $\sigma$
of a $k$-slab if $\sigma \in S$, and evaluates to $0$ otherwise. Let $S:=\{\sigma \colon w(\sigma) < w^*\}$, and 
let $T=\{ \sigma \colon f_\sigma \neq n\}$. Let
$ F = \sum_{\sigma \in \Re} \ind_S(\sigma) \ind_T(\sigma).$
Then the prover has supplied an optimal range $\sigma^*$ if and only if $F = |S|$. Note that effectively we are summing $\ind_T(\sigma)$ over a \emph{prefix} of the sorted list of ranges, namely those in $S$. 

Let $\field$ be a field of prime order satisfying $6 n^2 \leq |\field| \leq 6 n^3$. Let $\hat{\ind}_S \colon \field^{\log |\Re|} \rightarrow \field$ be the multilinear extension of $ \ind_S$, and let $\hat{\ind}_T$ be the multilinear extension of $\ind_T$. 
That is, $\hat{\ind}_S$ is the unique multilinear polynomial over $\field$ satisfying $\hat{\ind}_S(\sigma) = \ind_S(\sigma)$ for all $\sigma \in \{0, 1\}^{\log |\Re|}$, and 
similarly for $\hat{\ind}_T$.
It is standard that 
\begin{equation}
\label{mleeq} \hat{\ind}_S = \sum_{\sigma \in \{0, 1\}^{\log \Re}} \ind_S(\sigma) \cdot \chi_{\sigma}, \text{ where }\end{equation}
 \begin{equation}
\label{chieq} \chi_{\sigma}(x_1, \dots, x_{\log|\Re|}) := \prod_{i=1}^{\log|\Re|} (x_1 \sigma_i + (1-x_i)(1-\sigma_i)), \end{equation} 
and similarly for $\hat{\ind}_T$.
To compute $F$, it suffices to apply the sum-check protocol to the polynomial $g := \hat{\ind}_S  \cdot \hat{\ind}_T$. The protocol requires $\log |\Re|$ rounds,
and the total communication cost is $O(\log |\Re|)$ field elements. To perform the necessary check in the final round of this protocol,
$\textsf{V}$ needs to evaluate $g$ at a random point $\mathbf{r} \in \field^{\log |\Re|}$. By definition of $g$, it suffices for $\textsf{V}$ to evaluate $\hat{\ind}_T(\mathbf{r})$
and $\hat{\ind}_S(\mathbf{r})$.
Since the set $S$ does not depend on the stream ($S$ depends only on the claimed optimal width $w^*$),
$\textsf{V}$ can evaluate $\hat{\ind}_S(\mathbf{r})$ \emph{after} the stream has passed, using $O(\log(|\Re|) \cdot \log |\field|)$ bits of space, using standard
techniques (see for example \cite[Section 2]{cormode2011verifying}). However, it is not possible for $\textsf{V}$ 
to evaluate $\hat{\ind}_T(\mathbf{r})$ in a streaming manner. 
Instead, \textsf{V} asks $\textsf{P}$ to \emph{tell her} $\hat{\ind}_T(\mathbf{r})$, and checks that  
$\hat{\ind}_T(\mathbf{r})$ by invoking the streaming implementation of the GKR protocol (cf. Lemma \ref{lemma:muggles}). 
More precisely, similar to \cite[Section 3.3]{cormode2012practical}, we observe that Fermat's Little Theorem implies that 
$f_{\sigma} \neq n$ if and only if $(f_{\sigma}-n)^{|\field|-1} \equiv 1 \mod |\field|$. 
This implies via Equation \eqref{mleeq} that $\ind_T(\mathbf{r}) = \sum_{\sigma \in \{0, 1\}^{\log|\Re|}}(f_{\sigma}-n)^{|\field|-1} \cdot \chi_{\sigma}(\mathbf{r})$,
where $\chi_{\sigma}$ was defined in Equation \eqref{chieq}. 
As in \cite[Section 3.3]{cormode2012practical}, it is possible to compute the right hand side of this equality by a log-space uniform arithmetic circuit $\mathcal{C}$
of size $O(|\Re|)$ and depth $O(\log |\field|) = O(\log n)$ over $\field$. By applying the GKR protocol to $\mathcal{C}$,
$\textsf{V}$ forces $\textsf{P}$ to faithfully provide $\hat{\ind}_T(\mathbf{r})$. This completes the protocol. Completeness and soundness follow from
completeness and soundness of the sum-check protocol and of the GKR protocol. It is straightforward to check that the the protocol has the claimed space and communication costs.


\noindent \textbf{Protocol Costs.} 
The total communication cost of the protocol $O(\log n \cdot \log(| \Re|) \cdot \log |\field|) = O(k \cdot d^2 \cdot \log m \cdot \log^2 n)$  bits. 
The total space cost is $O(\log(|\Re|) \cdot \log(|\field|)) = O(k \cdot d^2 \cdot \log m \cdot \log n)$ bits.
The total number of rounds required is $O(\log n \cdot \log(| \Re|)) = O(k \cdot d^2 \cdot \log m  \cdot \log n)$.

\begin{theorem} \label{kslabtheorem}
Given a stream of $n$ points, there is a  streaming interactive proof for computing the optimal $k$-slab, 
with space and communication bounded by $O(k \cdot d^2 \cdot \log m \cdot \log^2 n)$. The total number of rounds is 
$O(k \cdot d^2 \cdot \log m \cdot \log n)$.
\end{theorem}
We note that it is possible to both avoid using the GKR protocol and reduce the number of rounds in Theorem \ref{kslabtheorem} by a factor of $\log(n)$, using a technique introduced by Gur and Raz \cite{gur}, and applied by Klauck and Prakash \cite{klauck2014improved}
to obtain an $O(\log |\Re|)$-round SIP for computing the number of distinct items in a data stream. However, these techniques sacrifice perfect completeness, and increase the communication complexity of the protocol by polylogarithmic factors. We omit the details of this technique for brevity.
\subsection{$k$-Center}
We can use the same idea as above to verify solutions for Euclidean $k$-center. The relevant range space here consists of unions of $k$ balls of radius $r$, for all choices of centers and radii in the grid. The size of this range space is $m^{2kd}$. We omit further details and merely state the main result. 
\begin{theorem}\label{k-center}
Given a stream of $n$ input points, there is an SIP for computing the optimal $k$-center with space and communication bounded 
by $O(k \cdot d \cdot \log m \cdot \log^2 n)$. The total number of rounds is 
$O(k \cdot d \cdot \log m \cdot \log n)$.
\end{theorem}



\bibliographystyle{splncs03}
\bibliography{survey}

\begin{thebibliography}{10}
\providecommand{\url}[1]{\texttt{#1}}
\providecommand{\urlprefix}{URL }

\bibitem{agarwal2010streaming}
Agarwal, P.K., Sharathkumar, R.: Streaming algorithms for extent problems in
  high dimensions. In: Proc 21st SODA. pp. 1481--1489 (2010)

\bibitem{andoni2013eigenvalues}
Andoni, A., Nguyen, H.: Eigenvalues of a matrix in the streaming model. In:
  Proc. 24th SODA. pp. 1729--1737 (2013)

\bibitem{babai1986complexity}
Babai, L., Frankl, P., Simon, J.: Complexity classes in communication
  complexity theory. In: {Proc. 27th IEEE FOCS}. pp. 337--347 (1986)

\bibitem{chakrabarti2013annotations}
Chakrabarti, A., Cormode, G., Goyal, N., Thaler, J.: Annotations for sparse
  data streams. In: SODA'14. pp. 687--706 (2014)

\bibitem{chakrabarti2009annotations}
Chakrabarti, A., Cormode, G., McGregor, A.: Annotations in data streams. In:
  Automata, Languages and Programming, pp. 222--234. Springer (2009)

\bibitem{chakrabarti2014annotations}
Chakrabarti, A., Cormode, G., McGregor, A., Thaler, J.: Annotations in data
  streams. ACM Transactions on Algorithms (TALG)  11(1), ~7 (2014)

\bibitem{chakrabartiinteractivity}
Chakrabarti, A., Cormode, G., McGregor, A., Thaler, J., Venkatasubramanian, S.:
  On interactivity in arthur-merlin communication and stream computation. In:
  Proc. IEEE Conference on Computational Complexity (2015)

\bibitem{chan06}
Chan, T.M.: Faster core-set constructions and data-stream algorithms in fixed
  dimensions. Computational Geometry  35(1),  20--35 (2006)

\bibitem{chung}
Chung, K.M., Kalai, Y.T., Vadhan, S.P.: Improved delegation of computation
  using fully homomorphic encryption. In: Rabin, T. (ed.) CRYPTO. Lecture Notes
  in Computer Science, vol. 6223, pp. 483--501. Springer (2010)

\bibitem{cormode2012practical}
Cormode, G., Mitzenmacher, M., Thaler, J.: Practical verified computation with
  streaming interactive proofs. In: Proc. 3rd ITCS. pp. 90--112. ACM (2012)

\bibitem{cormode2013streaming}
Cormode, G., Mitzenmacher, M., Thaler, J.: Streaming graph computations with a
  helpful advisor. Algorithmica  65(2),  409--442 (2013)

\bibitem{cormode2011verifying}
Cormode, G., Thaler, J., Yi, K.: Verifying computations with streaming
  interactive proofs. Proceedings of the VLDB Endowment  5(1),  25--36 (2011)

\bibitem{edwards2005no}
Edwards, M., Varadarajan, K.: No coreset, no cry: {II}. In: FSTTCS, pp.
  107--115. Springer (2005)

\bibitem{feder1988optimal}
Feder, T., Greene, D.: Optimal algorithms for approximate clustering. In: Proc.
  ACM {STOC}. pp. 434--444. ACM (1988)

\bibitem{DBLP:conf/soda/GoelIV01}
Goel, A., Indyk, P., Varadarajan, K.R.: Reductions among high dimensional
  proximity problems. In: Kosaraju, S.R. (ed.) Proc. 12th SODA. pp. 769--778
  (2001)

\bibitem{gkr}
Goldwasser, S., Kalai, Y.T., Rothblum, G.N.: Delegating computation:
  Interactive proofs for muggles. In: STOC '08. pp. 113--122. ACM, New York,
  NY, USA (2008), \url{http://doi.acm.org/10.1145/1374376.1374396}

\bibitem{gonzalez1985clustering}
Gonzalez, T.F.: Clustering to minimize the maximum intercluster distance.
  Theoretical Computer Science  38,  293--306 (1985)

\bibitem{gur}
Gur, T., Raz, R.: Arthur--merlin streaming complexity. Information and
  Computation  (2014)

\bibitem{nocore}
Har-Peled, S.: No coreset, no cry. In: Proc. {FSTTCS}. pp. 324--335 (2005)

\bibitem{klauck2011arthur}
Klauck, H.: On {A}rthur {M}erlin games in communication complexity. In: Proc.
  CCC. pp. 189--199 (2011)

\bibitem{klauck2014improved}
Klauck, H., Prakash, V.: An improved interactive streaming algorithm for the
  distinct elements problem. In: {Proc. ICALP}. pp. 919--930 (2014)

\bibitem{lfkn}
Lund, C., Fortnow, L., Karloff, H.J., Nisan, N.: Algebraic methods for
  interactive proof systems. J. {ACM}  39(4),  859--868 (1992)

\bibitem{mathoverflow}
O'Donnell, R.: Almost orthogonal vectors.
  \url{http://mathoverflow.net/questions/24864/almost-orthogonal-vectors/24886\#24886}

\bibitem{eurocrypt}
Papamanthou, C., Shi, E., Tamassia, R., Yi, K.: Streaming authenticated data
  structures. In: Johansson, T., Nguyen, P.Q. (eds.) EUROCRYPT. Lecture Notes
  in Computer Science, vol. 7881, pp. 353--370. Springer (2013)

\bibitem{vds}
Schr{\"o}der, D., Schr{\"o}der, H.: Verifiable data streaming. In: Yu, T.,
  Danezis, G., Gligor, V.D. (eds.) ACM Conference on Computer and
  Communications Security. pp. 953--964. ACM (2012)

\bibitem{thalersurvey}
Thaler, J.: Stream verification (2015), to appear in \emph{Encyclopedia of
  Algorithms}, Springer. Extended version available at CoRR, abs/1507.04188.

\end{thebibliography}

\newpage


\appendix

\end{document}